\documentclass[10pt]{article}

\usepackage{epsfig}

\newlength{\dinwidth}
\newlength{\dinmargin}
\marginparpush
5pt \topmargin -42pt \headheight 12pt

\newcommand{\fr}[2]{\left(\frac{#1}{#2}\right)}

\begin{document}
\titlepage

\vspace*{4cm}

\begin{center}

{\Large \bf On the distribution of collisionless particles in local potential well}

\vspace*{1cm}
\textsc{K. Belotsky\footnote{e-mail address: k-belotsky@yandex.ru}, 
M. Khlopov\footnote{e-mail address: Maxim.Khlopov@roma1.infn.it}} \\

\vspace*{0.5cm}
Moscow Engineering Physics Institute,
Moscow, Russia \\[0.5ex]
Center for Cosmoparticle Physics "Cosmion" of
Keldysh Institute of Applied Mathematics, \\
Moscow, Russia \\[0.5ex]
\end{center}

\vspace*{1cm}

\begin{abstract}
The distribution of collisionless particles with infinite motion in the presence of a local potential well is discussed. 
Such distribution is important for interpretation of results of dark matter searches. The relationship 
n/v=const, where n and v are respectively number density and velocity of particles, 
is derived for particles crossing a local potential well. The limits of application of this relationship are specified.
\end{abstract}

\vspace*{1cm}

The question on the distribution of collisionless particles inside 
the local potential well 
is important for the problem of Dark Matter (DM) searches (e.g., \cite{PS}-\cite{mon}). 
In the papers \cite{mon} a relationship for number density of DM particles $n$ inside local potential well 
\begin{equation}
\frac{n}{v}={\rm const},
\label{nv}
\end{equation}
where $v$ is the local velocity of DM particles in the reference frame related with the well, was used putting aside derivation.
We adduce derivation of Eq.(\ref{nv}) in the present note and specify the limits of its application 
to the problems of dark matter search.

In general, the particles which experience an infinite motion are of interest.
They are supposed to be distributed homogeneously 
at sufficiently large distance 
from the potential well.
Potential well is supposed to be spherically symmetric and small in size, 
so the particles cross it quickly and the effects of scattering can be neglected for the bulk of particles. 
These conditions, as a rule, are fulfilled for WIMPs crossing the Solar system.

Let us introduce necessary notations.
Intensity of the flux
\begin{equation}
I=\frac{dN}{dS_{n}dtd\Omega},
\label{I}
\end{equation}
where $N$ is the number of particles, $S_{n}$ is the area normal to given direction, $\Omega$ is the solid angle. 
If particles are distributed in velocity then $dv$ should be added in denominator of Eq.(\ref{I}).
In case of isotropic monochromatic distribution
\begin{equation}
I=\frac{nv}{4\pi}.
\label{Imon}
\end{equation}
The number of particles passing through given area $dS$ per unit time is
\begin{equation}
d\dot{N}=\int IdS\cos \theta d\Omega,
\label{flux}
\end{equation}
where $\theta$ is the angle between direction of incident particles and normal of area. In case of isotropic flux,
$\dot{N}$ from one side of area $dS$ ($0\le \theta\le \pi/2$) is
\begin{equation}
d\dot{N}=\pi I \cdot dS =\frac{nv}{4}dS.
\label{fluxis}
\end{equation}
For sphere of radius $R$ it gives the well known expression
\begin{equation}
\dot{N}_{on\,sphere}=\frac{nv}{4}\,4\pi R^2=nv\sigma,
\label{fluxis}
\end{equation}
where $\sigma=\pi R^2$ is the cross section of sphere.

Velocity of collisionless particles in a given point of potential well ($R$) can be defined through the relevant escape
velocity $v_{esc}(R)$
\begin{equation}
v_R =\sqrt{v_{\infty}^2+v_{esc}^2(R)}.
\label{vR}
\end{equation}
Throughout in the paper indexes "$R$" and "$\infty$" correspond, respectively, to the points 
at the distance $R$ and at "infinity"
(at sufficiently large distance, at which the influence of potential well is negligible).

Let us first consider a potential well of gravitating body with radius $R$ (e.g., a star). 
For particles moving from infinity the cross section of their capture by this body is given by
\begin{equation}
\sigma_{\infty}=\sigma_0 \left(1+\frac{v_{esc}^2}{v_{\infty}^2}\right)=\sigma_0 \frac{v_{R}^2}{v_{\infty}^2}.
\label{sinf}
\end{equation}
Here $\sigma_0=\pi R^2$ is the geometrical cross section of the body, $v_{esc}=v_{esc}(R)$ is the escape velocity on its surface,
and the Eq.(\ref{vR}) was used to derive the last expression in the Eq.(\ref{sinf}).
The number of particles, being isotropically distributed on infinity (though the following formula is not restricted by 
this case only), falling on star surface per unit time will be
\begin{equation}
\dot{N}=n_{\infty}v_{\infty}\sigma_{\infty}.
\label{Ninf}
\end{equation}
From the other hand the same number can be expressed through the corresponding values near the surface. 
If we suppose that distribution near the surface is also isotropic, we obtain
\begin{equation}
\dot{N}=n_Rv_R\sigma_0.
\label{NR}
\end{equation}
Equating Eqs.(\ref{Ninf},\ref{NR}) one gets
\begin{equation}
\frac{n_R}{v_R}=\frac{n_{\infty}}{v_{\infty}}.
\label{nv1}
\end{equation}
This derivation 
gives a hint for the Eq.(\ref{nv}), requiring to resolve the question of (an)isotropy inside potential well 
and to generalize the expression on the case of more realistic distribution.

Let us consider particles on infinity, having arbitrary angular distribution 
of their velocity vector ($\vec{v}_{\infty}$).
Assume that particles move in a central-symmetric potential field. The shape of potential well is not specified.
It will be enough in this case to use angular momentum conservation.

Let particles on infinity have the only value $v_{\infty}=|\vec{v}_{\infty}|$. 
One separates from them the flux concentrated along a given direction $I_{\infty}d\Omega$. Intensity can be factorized as
\begin{equation}
I_{\infty}=n_{\infty}v_{\infty}f_{\infty}(\theta,\phi),
\label{Ifact}
\end{equation}
where $f_{\infty}(\theta,\phi)$ is distribution function
in angle coordinates of $\vec{v}_{\infty}$ relatively to the centre 
of attraction normalized on unit
(e.g., for isotropic flux Eq.(\ref{Imon}) gives $f_{\infty}=1/(4\pi)$).

Impact parameter $\rho$, angle $\Psi$ of incidence on sphere of radius $R$ around gravitating center (equipotential layer)
are attributed to the motion of test particles (with mass $m$) of given flux as shown on Figure 1.
\begin{figure}
\begin{center}
\centerline{\epsfxsize=8.5cm\epsfbox{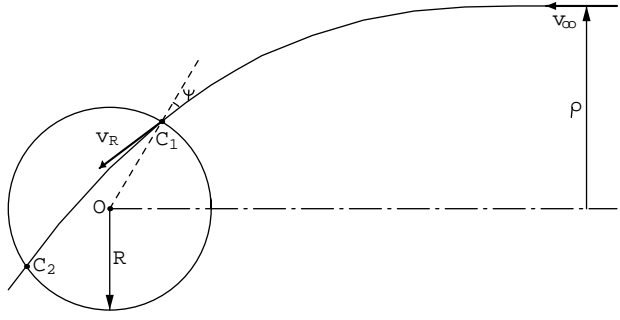}}
\caption{Illustration of motion of test particle in a central-symmetric attractive field with the center O.}
\label{fig}
\end{center}
\end{figure}

Angular momentum conservation law gives us
\begin{equation}
mv_{\infty}\rho=mv_R R\sin \Psi.
\label{angmom}
\end{equation}
Position where the particle crosses the sphere $C_1$ is not specified.
Note that
by force of Eq.(\ref{angmom}) an incident angle is simply defined by $R$ at the given $v_{\infty}\rho$.
In particular, an equality of the incident angles in the points $C_1$ and $C_2$ follows.
For the further use one can express $d\rho^2$ through $\cos\Psi$ (not caring about the sign to be eliminated by proper choice 
of integration limits)
\begin{equation}
d\rho^2=2 \fr{v_R}{v_{\infty}}^2R^2\cos\Psi d\cos\Psi.
\label{drho}
\end{equation}
The number of particles falling from the outside with the fixed impact parameter
$\rho...\rho+d\rho$ or corresponding incident angle $\Psi$ per unit time on sphere of radius $R$ is given by
\begin{equation}
d\dot{N}_{outside}=I_{\infty}d\Omega\cdot 2\pi\rho d\rho.
\label{dNdot1}
\end{equation}
For the number of particles crossing the given sphere both from the outside and from the inside
due to the above mentioned symmetry 
we have $d\dot{N}_{sph}=d\dot{N}_{outside}+d\dot{N}_{inside}=2d\dot{N}_{outside}$. 
Transforming it with the use of Eq.(\ref{drho}) one obtains
\begin{equation}
d\dot{N}_{sph}=I_{\infty}d\Omega\cdot 4\pi R^2 \fr{v_R}{v_{\infty}}^2 \cos\Psi d\cos\Psi.
\label{dNdot2}
\end{equation}
One can notice from the comparison of Eq.(\ref{dNdot2}) and Eq.(\ref{flux}) that intensity of particles falling on sphere
(or near it) integrated over asimuth angle of $\vec{v}_R$ (where angular coordinates are
defined relative to $\vec{R}$) is $dI_R(\Psi)=I_{\infty}d\Omega \cdot \fr{v_R}{v_{\infty}}^2={\rm const}$.

Consider particles within spherical layer of radius $R$. Let us calculate the number of particles present
at the same time inside the layer $R...R+dR$. Each particle moving through the layer with incident angle $\Psi$ will
reside in it during time
\begin{equation}
d\tau=\frac{dR}{v_R\cos\Psi}.
\end{equation}
Then the mean number of particles present in the layer will be
\begin{equation}
dN_{layer}=d\dot{N}_{sph}d\tau=I_{\infty}d\Omega\cdot \frac{4\pi R^2dR}{v_R} \fr{v_R}{v_{\infty}}^2 d\cos\Psi.
\label{Ninlayer}
\end{equation}
Note
that the distribution on $\cos\Psi$ of number of particles 
(of their arbitrary flux from infinity) inside the layer is constant, i.e.
\begin{equation}
\frac{dN_{layer}}{d\cos\Psi}={\rm const}.
\label{Ncos}
\end{equation}
It is worth to emphasize that the angle $\Psi$ is defined
relative to the direction towards the center of attraction (O) $\vec{R}$ wherever the point $C_1$ is situated on the sphere.
It does not imply
space homogeneity of distribution of the points $C_1$ on sphere. 

After trivial 
integration of Eq.(\ref{Ninlayer}) over $\cos\Psi$ in the limits from 0 to 1 
($\cos\Psi=0$ corresponds to maximal $\rho=\frac{v_RR}{v_{\infty}}$)
one writes with the use of Eq.(\ref{Ifact})
\begin{equation}
\frac{dN_{layer}}{4\pi R^2dR}=\frac{n_{\infty}}{v_{\infty}}v_R\cdot f_{\infty}(\theta,\phi)d\Omega.
\label{Ninlayer1}
\end{equation}
The value $\frac{dN_{layer}}{4\pi R^2dR}$ is the number density of particles averaged over sphere - 
equipotential layer, $\bar{n}_R$. After integration over $d\Omega$ - directions of incident particles from infinity,
one obtains
\begin{equation}
\frac{\bar{n}_R}{v_R}=\frac{n_{\infty}}{v_{\infty}}.
\label{nv2}
\end{equation}

In case of isotropic distribution $f_{\infty}$ at infinity one can generalize Eq.(\ref{Ncos}) as
an averment of isotropy of incident flux locally in any point of (any) equipotential layer. 
Eq.(\ref{nv2}) in this case takes the form of 
Eqs.(\ref{nv1}),(\ref{nv}) (i.e. averaging over the layer is not necessary).

To generalize given above arguments for a case of arbitrary distribution in velocity $v_{\infty}$,
one introduces $dv$ in denominator of Eq.(\ref{I}). One separates the flux 
$I_{\infty}dv_{\infty}d\Omega=n_{\infty}v_{\infty}f_{\infty}(v_{\infty},\theta,\phi)dv_{\infty}d\Omega$. 
Then Eq.(\ref{Ninlayer1}) takes the form
\begin{equation}
d\bar{n}_R=\frac{n_{\infty}}{v_{\infty}}v_R\cdot f_{\infty}(v_{\infty,}\theta,\phi)dv_{\infty}d\Omega.
\label{Ninlayer2}
\end{equation}
For any point $i$, a notation $dn_i=n_if_i(v_i,\theta,\phi)dv_id\Omega=n_if_i(v_i)dv_i$, 
where indexes for angles are omitted and $\int f_i(v_i,\theta,\phi)d\Omega=f_i(v_i)$, is adopted 
($n_i$ and $f_i$ are overlined, if any).
Velocities $v_{\infty}$ and $v_R$ are mutually related 
by Eq.(\ref{vR}) and for their differentials the relationship
\begin{equation}
v_Rdv_R=v_{\infty}dv_{\infty}
\end{equation}
is valid.
Integration of Eq.(\ref{Ninlayer2}) over $d\Omega$ with introduced notation yields
\begin{equation}
\frac{d\bar{n}_R}{v_R}=\frac{dn_{\infty}}{v_{\infty}}.
\label{nv3}
\end{equation}
So, account of distribution in velocity leads to replacement $n_i \rightarrow dn_i(v_i)$ in the Eq.(\ref{nv2}).

Velocity distribution, averaged over sphere of radius $R$, is determined through that on infinity as
\begin{equation}
\bar{f}_R(v_R)=\frac{n_{\infty}}{\bar{n}_R}\frac{v_{R}^2}{v_{\infty}^2(v_R)}f_{\infty}(v_{\infty}(v_R)).
\end{equation}
Its normalization on unit is provided by factor $\frac{n_{\infty}}{\bar{n}_R}$. Note, that $\bar{f}_R(v_R)$ has
a threshold $v_R\ge v_{esc}(R)$, where it experiences a jump from zero.

The relations for number density considered above
are applied to the particles with infinite motion and
do not take into account particles captured by potential well. The latter would contribute in range just below 
the threshold $v_{esc}(R)$ (and in perfect case would reproduce equilibrium form of velocity distribution).
In case of WIMPs trapped by Solar system, still there is no definite answer 
on the effectiveness of respective mechanisms of 
the capture and, especially, of the escape of WIMPs (see, e.g., \cite{LE}, \cite{DK}, \cite{G3}).

Application of Eq.(\ref{nv}) or Eq.(\ref{nv3}) for estimation of accumulation of WIMPs by the Earth 
can cause errors. The Earth is inside Solar potential, 
what could be taken into account with "two-step" application of given equation.
One can get first the proper WIMPs parameters inside the potential of the Sun at the distance from it corresponding 
to the Earth orbit
and then pass to the Earth reference frame and get those parameters inside the Earth potential. However,
the use of Eq.(\ref{nv3}) at first step will give density distribution averaged over all equipotential layer around the Sun 
rather then that along the Earth's orbit only. 

Because of the averaged form of the obtained result Eq.(\ref{nv}) can not be applied to the analysis of annual modulation
in direct WIMP searches \cite{DAMA}.

One of the implications of given here equations can be connected with predictions of possible existence of anti-meteorits 
in our Galaxy \cite{FK}. Analysis of the effects of their annihilation on Sun can provide crucial test
of this prediction. For estimation of frequency of such events Eq.(\ref{nv}) can be used and
to study observable effects (specific forms of flares) produced by them Eq.(\ref{Ncos}) can be useful.

Generalization of Eq.(\ref{nv}) on case of $D$-dimensional space has the form
\begin{equation}
\frac{n}{v^{D-2}}={\rm const}.
\end{equation}
In case $D=1$ it leads to the familiar result from theory of gas flow in tube with variable cross section.

\begin{center} {\bf Acknowledgment}\end{center}
K.B. would like to express his gratitude for a help to V.I. Kogan and Yu.D. Fiveyskiy, rendered in its time. 
M.Kh. is grateful to CRTBT-CNRS, Grenoble for hospitality.
The work was supported in part by grant of Khalatnikov-Starobinsky scientific school.

\end{document}